\documentclass{article}

\newcommand{\mycomment}[1]{}
\usepackage{arxiv}

\usepackage[utf8]{inputenc} 
\usepackage[T1]{fontenc}    
    
\usepackage{url}            
\usepackage{booktabs}       
\usepackage{amsfonts}       
\usepackage{nicefrac}       
\usepackage{microtype}      
\usepackage{lipsum}
\usepackage{graphicx}
\graphicspath{ {./images/} }

\usepackage[style=numeric-comp, sorting=none, backend=biber]{biblatex}
\usepackage[colorlinks=true, allcolors=blue]{hyperref}

\author{
  Alexander Michael Rusnak \\
  Digital Humanities Laboratory \\
  École Polytechnique Fédérale de Lausanne 
  }

\title{Representing Beauty: Towards a Participatory but Objective Latent Aesthetics}

\begin{document}
\maketitle
\begin{abstract}
What does it mean for a machine to recognize beauty? While beauty remains a culturally and experientially compelling but philosophically elusive concept, deep learning systems increasingly appear capable of modeling aesthetic judgment. In this paper, we explore the capacity of neural networks to represent beauty despite the immense formal diversity of objects for which the term applies. By drawing on recent work on cross-model representational convergence, we show how aesthetic content produces more similar and aligned representations between models which have been trained on distinct data and modalities - while unaesthetic images do not produce more aligned representations. This finding implies that the formal structure of beautiful images has a realist basis - rather than only as a reflection of socially constructed values. Furthermore, we propose that these realist representations exist because of a joint grounding of aesthetic form in physical and cultural substance. We argue that human perceptual and creative acts play a central role in shaping these the latent spaces of deep learning systems, but that a realist basis for aesthetics shows that machines are not mere creative parrots but can produce novel creative insights from the unique vantage point of scale. Our findings suggest that human-machine co-creation is not merely possible, but foundational - with beauty serving as a teleological attractor in both cultural production and machine perception.
\end{abstract}


\section{Representation in Art and Computer Science}

Representation is a central concern of both art and deep learning. Influential Cubist critic Jacques Rivière asserts: ``The true purpose of a painting is to represent objects as they really are; that is to say, differently from the way we see them. It always tends to give us its sensible essence, its presence, this is why the image it forms does not resemble its appearance...” \cite{riviere1992present}. \mycomment{A common definition of the artistic act is as an instantiation of thoughts, intentions, beliefs, and desires into some form such as a painting or a performance. Direct pictorial representation is prevalent, but so are more abstract forms of representation which seek to depict partially or wholly immaterial concepts. Similarly, though often with radically different tools, representation learning in computer science attempts to take input data (for example, images) relative to a particular domain or concept (such as flight or birds) and distill them into a numerical form which encapsulates the informational essence of the data;} Similarly, the authors of the influential infoGAN paper \cite{NIPS2016_7c9d0b1f} state that the goal of representation learning ``is to use unlabeled data to learn a representation that exposes important semantic features as easily decodable factor”. In both contexts, the semantic representation of physical objects is more straightforward: an apple possesses many relatively concrete properties (such as color or size) which emerge easily from observation, either statistically or phenomenologically. In contrast, attempting to represent more abstract concepts is inherently more challenging due to the diversity of objects or situations for which the concept can apply. Extending the furthest towards the abstract there are transcendental concepts such as beauty - which exists as a phenomenological experience and, as we argue, has a physical basis despite the term being applied with massive formal variety to mountains, sunsets, paintings, people, acts of kindness, and so forth. The inherent tension between the semi-mysterious, conceptually diverse instantiations of beauty and its perceptual centrality in the human experience has made representing it through various forms a key concern of artists, authors, musicians, and creative professionals throughout history. Furthermore, despite assertions of the total  subjectivity and cultural construction of aesthetics by critics from the postmodern\cite{postmodern-aesthetics}, pragmatist\cite{dangelo2012pragmatist}, and anti-realist camps\cite{Andow04072022}, it has been demonstrated that deep learning systems\cite{aestheics-survey} - even those trained on more general tasks\cite{ conwell2024usingmultimodaldeepneural, clip-knows} - have been able to accurately predict image aesthetics across diverse benchmarks, giving credence to realist or at least strongly inter-subjective conceptions of beauty. The changing artistic landscape caused by the emergence of deep learning based tools has also made digitally mediated representation a crucial topic to explore\cite{schaerf2024reflectionsdisentanglementlatentspace}. Following these observations, we believe beauty merits a closer look through the lens of representation learning. 

\begin{figure*}

\begin{center}

  \includegraphics[width=1\linewidth]{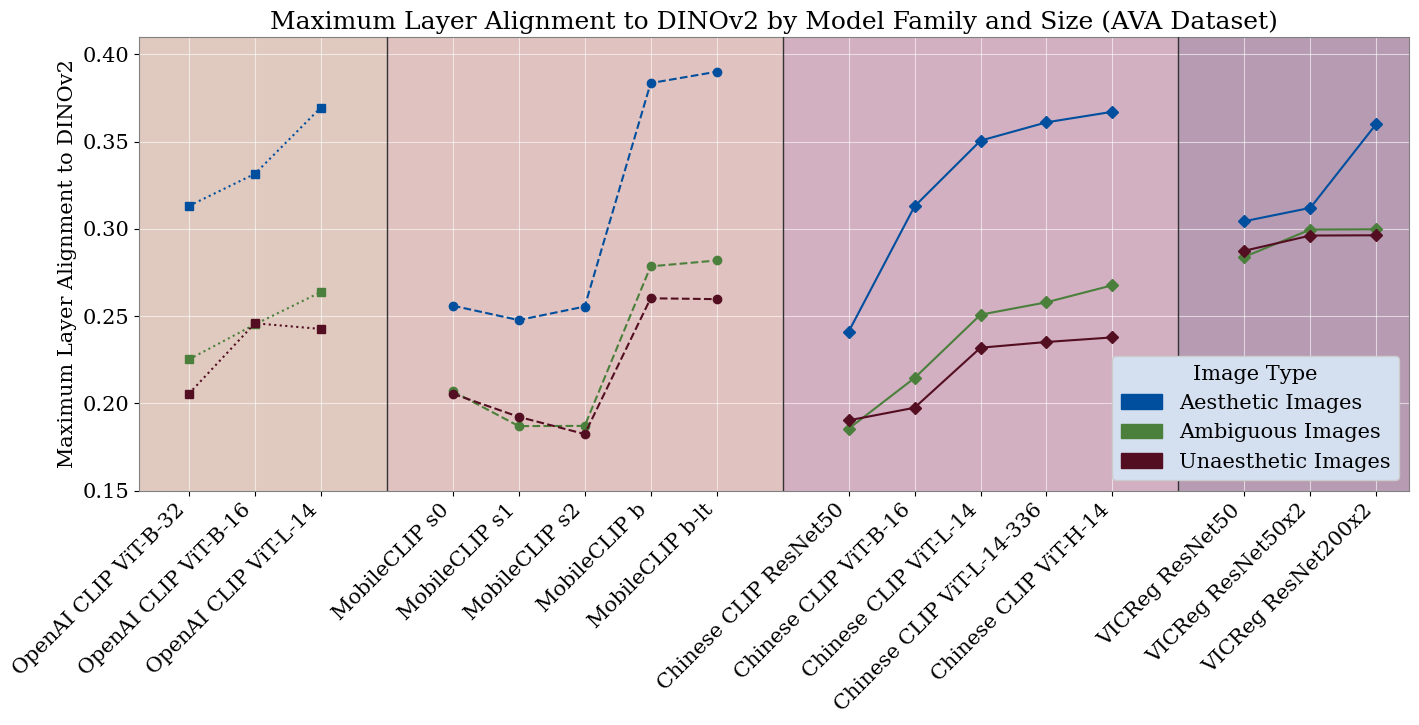}

\end{center}

   \caption{The cross-model mutual nearest neighbors representational alignment (following Huh et al.\cite{platonic}) between DINOv2-Large \cite{oquab2024dinov2learningrobustvisual} and a variety of models (CLIP \cite{radford2021learningtransferablevisualmodels}, MobileCLIP \cite{vasu2024mobileclipfastimagetextmodels}, ChineseCLIP \cite{yang2023chineseclipcontrastivevisionlanguage}, and the fully self-supervised and unimodal VICReg \cite{bardes2022vicregvarianceinvariancecovarianceregularizationselfsupervised}) with diverse training data, architectures, and training schemes. The representational alignment of each model is  delineated by the aesthetic classification of the source images, all of which are derived from the Aesthetic Visual Analysis dataset \cite{AVA}. There is a distinctly higher level of representational alignment between the aesthetic images than between the unaesthetic images or the representations corresponding to aesthetically ambiguous images.}

\label{fig:intra-inter}

\end{figure*}

Additional intrigue is added to the discussion of abstraction and representation by the ongoing debate regarding the ``Platonic representation hypothesis” (PRH), which states ``that there is a growing similarity in how datapoints are represented in different neural network models. This similarity spans across different model architectures, training objectives, and even data modalities” and ``that there is indeed an endpoint to this convergence and a principle that drives it: different models are all trying to arrive at a representation of reality, meaning a representation of the joint distribution over events in the world that generate the data we observe” \cite{platonic}. In other words: that there is not an endless subjectivity of possible representations which are agnostic to each other, but that there is indeed a globally maximal (universal) representational geometry, regardless of the modality by which particular concepts or objects are presented to a neural network or the human cognitive system. Though various forms of this hypothesis have been prevalent in the history of philosophy and computer science, the authors support their claim by presenting evidence that representations produced by different models with distinct, non-overlapping modalities and datasets converge as the models become more successful across multiple benchmarks and thus more able to accurately apprehend reality. Following the original PRH paper, numerous other researchers have replicated and expanded on their approach, finding  agreement, with some small caveats, on the key observation of increasing cross-model representational convergence being driven by increased model performance on various tasks. In terms of bolstering the underlying claims of representational convergence through semantic abstraction, we refer to only a few of the available papers: Wei et al. \cite{wei2022emergentabilitieslargelanguage}, Vafidis et al. \cite{vafidis2025disentangling}, Jha et al.\cite{jha2025harnessinguniversalgeometryembeddings}, and Ramidi et al.\cite{ramidi2026representationalalignmenthypothesisevidence}. Wei et al. substantiates that the abilities of LLMs, and by extension other deep learning models, advances not as a linear function of training data, but coinciding with the emergence of abstractions which enable new forms of reasoning (such as a grasp of analogy or sarcasm) - something supported by more formalized research like Yang et al. \cite{yang2025emergent}. Vafidis ``proved that disentangled, generalizable representations must emerge in agents
optimally solving multi-task evidence accumulation tasks” - a key claim underlying the convergence of representational geometry. Jha et al. demonstrated that unpaired translations between embedding spaces were possible, a phenomenon they deem the \textit{strong Platonic representation hypothesis} because it implies not just local neighborhood alignment but sufficient mutual information in representational geometry as to make the embeddings translatable across embedding spaces with no paired data or language supervision. Lastly, Ramidi et al. presents a compelling survey of various techniques, as well as a delineation of sub-claims relevant to representational alignment, before eventually concluding: ``Taken together, these ... lines of evidence suggest that seemingly disparate modalities, trained under different objectives and data streams, frequently exhibit a coherent semantic geometry that can be revealed either by learning a simple mapping or by comparing their internal relational patterns.”

As a last volley against claims that representational alignment must be the result of common / similar datasets or network architectures, rather than being driven by abstraction based on the realist conceptual relations between physical objects, we present evidence from neuroscience showing that deep learning models and the human brain likewise converge to similar representational geometry: Gao et al. \cite{Gao2025}, Goldstein et al. \cite{Goldstein2025}, and Lopez et al. \cite{lopez-cardona2025brainlanguage}. 

It is clear from the density of attention given to this viewpoint by computer scientists as well as the repeated replication of results across varying domains or research teams that representational alignment cannot be easily dismissed by humanities scholars because it may conflict with fashionable theories about the structure of knowledge. 

\mycomment{ If this hypothesis were true, we would also expect there to be some learnable content within the embeddings produced by various models such that we can translate embeddings between architectures without relying on labels or pairs, but instead on their shared geometry. And this is exactly what Jha et al.\cite{jha2025harnessinguniversalgeometryembeddings} found: that it is possible to translate (albeit imperfectly) between embedding spaces by constructing a bridge through a shared, quasi-universal latent space. }

In this short paper, primarily framed through the philosophy of art, we will make a few key assertions/agreements in order to contribute to debates around the creative usage of artificial intelligence, aesthetics, and the source/structure of potentially universal embeddings:

\begin{itemize}

    \item  We echo the claim that the human cognitive system creates some internal representations of the observed world in order to understand and act within it\cite{rudrauf2023projective, platonic, CAO2024101200, luo2019neural}, which are constrained in their structure by the physical world and shared biology.
    \item We claim that these representations, in humans or as recognized by machines, are more accurately categorized as hylomorphic (in the Aristotelian sense) than Platonic because of they are closely coupled with and derivative of the physical world.
    \item We claim that transcendent ideals or concepts, beauty being the most relevant in the case of art, act as ``binders” for this universal latent space due to their high relational centrality to other concepts and their teleological significance to human beings. 
    \item We claim that humans project these representations into the world through acts of ``creation” - including using language or creating art, but also through more mundane tasks such as capturing or captioning images. 
    \item We claim that this human projection is of key importance to the ability of deep learning systems to efficiently model the world by bootstrapping a quicker path to apprehending universal representations. 
\end{itemize}

We take a deliberate position against dominant constructivist and relativist theories of latent aesthetics, exemplified by Impett's modified quotation of Anthony Giddens, used in reference to the AVA dataset to claim its expression of aesthetics is purely socially constructed: ``Machine Learning is not about a `pre-given' universe of objects, the universe is being constituted — or produced by — the
active doings of subjects” \cite{impett2019robot, giddens1993new}. We instead argue for an objective, but human-mediated, structure to beauty as evidenced by the internal coherence of aesthetic embeddings in large-scale deep learning models. We contend this framing opens new possibilities for interpreting AI systems not as mere mimics, but as devices for uncovering latent order in cultural and physical reality, which should change the way artists use these tools. We draw on a lineage that treats beauty not as constructed taste but as a manifestation of order and harmony; within the art world this lineage includes classical, formalist, and essentialist movements that approach artistic creation as a search for universal structure or values. In this context, neural networks trained on large corpora of culturally filtered data may serve as uniquely sensitive instruments for detecting such structure.

For the remainder of this paper, we will utilize the term ``universal representation hypothesis” to refer to this set of ideas and findings around the possibility of a universal geometry, (except when referring directly to the work from Huh et al.) because part of our contention is that ``Platonic” is a misleading characterization of the structure and source of these representations. 

\subsection{Contravariance and Contrapposto}

In order to properly contextualize the significance of these universal representations for creative understanding and action, it is crucial to understand their source. Following Cao et al.\cite{CAO2024101200} and Huh et al.\cite{platonic}, we cite task-based and architecture-based constraints as drivers for creating consistent representational geometry across systems. In regards to the task-based constraints that produce models, biological or otherwise, with particular forms (and thus representations produced by these systems), Cao et al. state: ``Why does the system have this form? Answer: Because it had to perform that function – and, importantly, the world is such that having to perform a particular function rules out many other forms.” Cognitive and computational architectures are subject to these constraints in order to solve for various functions, but they likewise exhibit their own intrinsic pressure on both future iterations of an architecture and the representations produced by a particular system. In the context of deep learning, any architecture constrains the possible representations it produces in trivial ways such as the hidden dimension / embedding size or in the flexibility of a system relative to the number of parameters, as well as in non-trivial ways such as the topology of the embedding space\cite{b2024topologicalperspectivesoptimalmultimodal} or its intrinsic dimension\cite{tsukagoshi-sasano-2025-redundancy}.  \mycomment{The success of particular architectures also constrains the future iterations of a system by placing a cost on innovation - not only does strong investment into a particular subs-system or component in some way bake it into the larger system by making it more expensive to replace, but in some cases forces a situation where poorer performance in the short term must be paid to retreat into the correct branch of a developing technological tree. This means that derivative systems are more likely to be similar than in a theoretical situation where there is no penalty for radical changes to a problem solution. } In discussing the ability for artificial systems to replicate brain structure and ability, Cao et al. define a relationship between problem complexity its realizable solutions: ``Though it may at first seem counterintuitive, the harder the computational goal, the easier the model-to-brain matching problem is likely to be. This because the set of architectural solutions
to an easy goal is large, while the set of solutions to a challenging goal is comparatively smaller.
In mathematical terms, the size of the set of optima or solutions is \textit{contravariant} in the difficulty
of the optimization problem.” In humans, they primarily discuss the genomic component of evolution as an optimization mechanism because their main concern is neurological modeling, while leaving cultural evolution primarily untouched. Yet changing technology, memetic devices, and emergent cultural perspectives are a substantive avenue for introducing low-cost variance as an vehicle for optimization\cite{richerson2008cultural_evo}. Within the cultural evolutionary landscape, art is a primary method of frontier innovation and communication of complex ideas or sentiment\cite{evo_art1}.  

As a direct metaphor for the contravariance principle, consider the emergence of \textit{contrapposto} in Renaissance sculpture, exemplified by Michelangelo’s \textit{David}. Its dynamic pose was constrained by both physical limits (the strength of marble, the balance of weight), cultural ideals of naturalistic beauty, and Michelangelo's own \textit{concetto} of the good and the beautiful\cite{howard2023giovio, agoston1997sonnet, nims1998four}. These constraints co-shaped the final form: the tree stump added for structural stability is more a result of material necessity than the aesthetic intention which shaped the proportions of \textit{David}. In this way, the sculpture emerges not from abstraction alone, but from a convergence of physical, perceptual, and teleological constraints.

\mycomment{
Further examining the subject of art, we can look to the emergence of the and usage of the contrapposto pose as a general demonstration of the contravariance principle in action both materially and culturally. Contrapposto, as perhaps best exemplified by Michelangelo's \textit{David}, arose out of a desire to create more realistic sculptures whose poses are not as static and lifeless as their flat-footed counterparts. The possible forms of this sculptural style are obviously constrained by the limits of the human body, but also by the material of rendering. Rendering this pose in marble is fragile because it places the majority of the weight of the sculpture onto one leg. This balance forces an adaptation by necessitating bracing structures, such as a tree stump in the case of \textit{David}, substantially modifying the total form of the sculpture. So the form of the final sculpture was optimized out of a combination of material constraints (i.e. the strength of marble), naturalistic constraints (mimicry of a natural pose held by a natural looking man), and Michelangelo's own \textit{concetto} of the beautiful\cite{howard2023giovio, agoston1997sonnet, nims1998four}. Upon further inspection, the material and naturalistic constraints are also aesthetic in origin: the choice of marble and desire for rendering a realist sculpture both stem from Michelangelo's, and the surrounding culture's, conception of the good and the beautiful. 

}

\subsection{Hylomorphism and Beauty}

Though we earlier utilized the Platonic language of Huh et al.\cite{platonic} to describe their findings, we assert that a Platonist philosophical framing of this universal representation hypothesis is insufficiently descriptive. The underlying Platonic conception they reference is of a pure ``world of ideal forms” which exists outside of the physical world and projects form and semantics into an impure physical reality. However, the authors offer the `contravariance principle' as an explanation for the representations recovered by these large models which is directionally the opposite relationship: the ``ideal” representations are existent but somehow implicit in the physical world - not only present in a detached 
metaphysical realm - and are learned from the constraints produced by a natural environment. This maps far more cleanly onto an Aristotelian conception of form and matter, hylomorphism: that every physical object is a unified substance of matter (\textit{hulê}) and form (\textit{morphê})\cite{ainsworth2024form_stanford, charles2023history}. From this perspective, form is an implicit order which is coterminous with matter and is not distinct from the physical world. Hylomorphism is a far more apt description of the logic given to explain converging representations than a neo-Platonist conception of form because it recognizes the indispensability of the physical world. Other authors have agreed an Aristotelian framing is more apt, though in their case primarily for his theory of relatives rather than Hylomorphism \cite{groger2026revisiting}.

Building on this philosophical grounding, we can reason around the importance of beauty in understanding the underlying logic of form. Following Kant, we can reiterate that for the world to be intelligible for us or machines, there must be some underlying order (systematicity) and ``to presuppose in an a priori and subjectively justified fashion that the world is systematically ordered under hierarchies of laws such that it has ‘a regard to our faculty of cognition’. Elsewhere Kant says that this presupposition is downright required for us rationally to engage in scientific enquiry, and perhaps even to form any empirical concepts whatsoever”\cite{Chignell2006-beauty-system}. Though this systematicity is diffusely realized in the physical world, in its totality it is only understandable as a supersensible, transcendental idea; that is one whose full structure is beyond empirical measurement and thus must be reasoned over through the use of analogy and symbology. But what symbol or symbols would be appropriate for a unifying order which connects and structures a diverse and chaotic world? Following the Scholastics\cite{pasnau2024aquinas, eco1988aesthetics} and Leibnez\cite{leibnez}, Kant assumes ``a common characteristic of the objects that we judge to be beautiful is that they contain a multitude of different shapes, sounds, tastes, and structures which are unified in a harmonious, organic fashion: flowers, fantasias, crustaceans, birdsongs, curlicues on wallpaper, and so forth.” He goes on to argue that beauty is symbolic of natural systematicity because ``such beautiful things would then be structurally analogous to a world-whole that is diverse and maximally specific and yet harmoniously ordered under a hierarchical system of natural laws” \cite{Chignell2006-beauty-system}. Circling back to the topic of the universal geometry of embeddings, we can see that beauty is likewise symbolic of or analogous to this geometry because it too orders and unites diverse concepts and modalities under a shared structure without destroying their particularities. Furthermore, the inhuman amount and diversity of data which can be processed by machines may open paths towards new insights about or analogies for these transcendent concepts.

How exactly is beauty analogous or descriptive of universal representations? There is a crucial distinction between extrinsic and intrinsic analogy; ``analogies of extrinsic attribution are ‘improper’: they are a manner of speaking, the application of a name where, ultimately, it does not properly belong”, whereas ``with an intrinsic use ‘there exists one common formal and objective concept’ across the varied uses of the word”, ``we could say that
not only Shakespeare himself, but also his plays, can be intrinsically Shakespearean” \cite{analogy}. In our case, beauty is in an important conceptual category in the sense that it can be applied intrinsically, that is appropriately, across a large manner of highly diverse objects - ``buildings and ballet can both be called beautiful because both are beautiful ... once analogy is intrinsic, the
linguistic aspect serves more to mark something than to make it so” \cite{analogy}. This `linguistic marking' or symbol points towards a shared material basis for the emergence of the ability to perceive beauty in both humans and machines - the same natural systematicity theorized by Kant which underlies the intelligibility of reality. If beauty was applied only as extrinsic analogy to the physical world, then we would not expect deep learning systems without access to human analogy (i.e. unimodal image models trained with self-supervised learning in our case) to be able to capture the formalist structure of beauty either as individual representations or the global representational geometry. \mycomment{Especially because many deep learning systems which have not been explicitly designed or encouraged to understand the beautiful still develop representations for it - following the contravariance principle outlined earlier, this should make us more certain that some other constraint regarding aesthetics has influence on the models, whether their source is natural systematicity directly or through mapping to human phenomenological descriptions by indirect labels of beauty.}

\section{The Teleological Hierarchy of Abstraction: From Particular to Transcendent}
\label{gen_inst}

\mycomment{Though we find a structural analogy between the natural ordering of beauty and embedding spaces, we hypothesize that there is a more direct relevance to transcendental concepts in a properly structured embedding space.} Beyond mere structural analogy, we hypothesize a direct relevance of transcendental concepts to structured embedding spaces. Representation learning works by processing many particulars (such as images of unique, distinct chairs with \textit{haecceitas}\cite{scotus1891opus, eco1988aesthetics, sep-medieval-haecceity}) through abstraction to find general representations (the concept or ideal of a chair). To illustrate this, we can once again draw an analogy to art - specifically modernist abstraction. Constantin Brancusi, arguably the most significant modernist sculptor, said: ``What is real is not the external form, but the essence of things... it is impossible for anyone to express anything essentially real by imitating its exterior surface” \cite{read1964concise}. Brancusi's sculpture \textit{L'Oiseau dans l'espace} demonstrates this impulse clearly; in an attempt to represent the essence of a bird, he produced an abstracted, minimalist sculpture with little direct pictoral relationship to any particular bird. This same process of abstraction appears to be at work when it comes to the representation of transcendentals such as beauty - in order to recognize the beauty of a particular object, we articulate its form in relation to various more abstract categories of relevance and also as a unified object or visage. In order to apprehend beauty, we reason or perceive at a higher level of abstraction over the ideals of many diverse objects for which the term is appropriate - because of the formal, aesthetic, and conceptual diversity of the beautiful this deep level of abstraction must be possible to make the concept intelligible. We uncover a hierarchy of abstraction from particular to general / universal, and from general / universal to the transcendent. 

\begin{figure*}[!ht]
\begin{center}
  \includegraphics[width=0.49\linewidth]{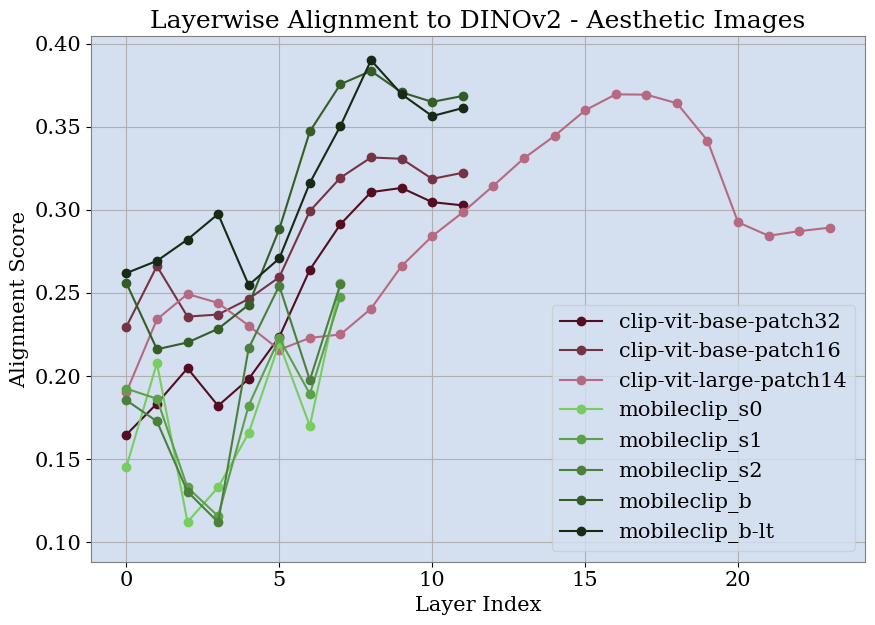}
  \includegraphics[width=0.49\linewidth]{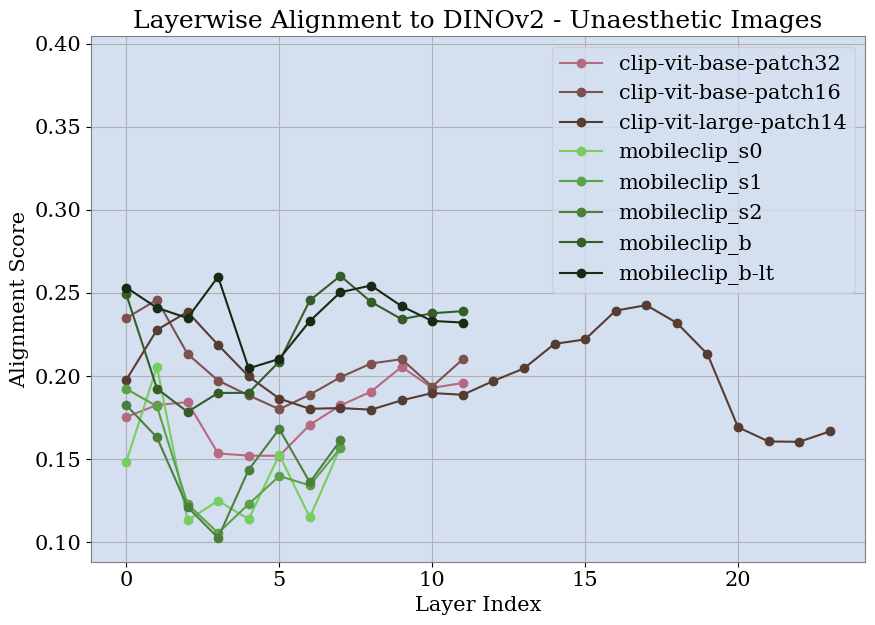}
\end{center}

   \caption{The layerwise alignment to DINOv2-Large for multiple models and corresponding to aesthetic and unaesthetic images. Not only do the aesthetic representations have higher overall alignment, but they also demonstrate the abstraction paradigm we described more clearly, where the middle layers display more universal, abstract representations.}
\label{fig:layer-align}

\end{figure*}

This conception of hierarchical abstraction in representation is complicated by the enforced flatness in the most common embedding models designed for retrieval (such as CLIP). A flat conceptual structure is more efficient for querying and thus trained towards - but in a sense all neural networks capture intermediate representations that distinguish various concepts in different topological arrangements and which could be articulated as a layer-wise hierarchy. In fact, there is some evidence of exactly this occurring within larger models;\cite{lindsey2025biology} Lindsey et al. and Yang et al.\cite{yang2025emergentsymbolicmechanismssupport}, report abstraction mechanisms at work in language models which transform particular tokens into a symbolic space, performing operations such as collocating words in different languages by their semantic meaning or by their inferred causal relations to other symbolic variables. Additionally, Skean et al.\cite{skean2025layerbylayer} and Bolya et al.\cite{bolya2025perceptionencoderbestvisual} both argue that optimal features for particular tasks exist at earlier layers of the network before the final embeddings, showcasing that optimizing for a traditionally regulated CLIP-like embedding space can suppress other real and informative features. All of these results are evidence for the particular to general type of abstraction, and we show in Figure \ref{fig:layer-align} that aesthetic content produces more aligned internal layers to partially substantiate the general to transcendent form of abstraction. 

Assuming this hierarchical abstraction mechanism is the case based on the preliminary but still compelling evidence, what sort of task based causal drivers might be forcing this learning? On one hand, explicit aesthetic evaluation of content is an important task in its own right for artists, designers, musicians, and other creatives, but we argue that implicit aesthetic evaluation is inherent to the human production and dissemination of digital content. As a concrete example, the photos which are captured and then curated to the point of being uploaded to the internet, even by the cultural laity, are still largely filtered by aesthetic quality; many of those which are reduplicated or spread broadly often even more so by the independent evaluation of the viewers. Most people have a conception of craft within their field of work and thus in orienting their work towards quality implicitly align it towards beauty as well. In aggregate, this creates a natural telos within human culture towards beauty which acts implicitly to curate content produced by humans. To understand how this could effectively influence the representations produced by deep learning systems, or recursively act on the human cognitive system\cite{luo2019neural} through cultural co-evolution as previously discussed, it is helpful to return to Cao et al.\cite{CAO2024101200}, who summarize that developed cognitive systems ``point at something like a normative function: a task to be performed, or a goal to be achieved. In each case, the system — the organism, the brain, or the neural network abstraction — is \textit{teleonomic}; it was shaped — by evolution, development, learning or training — \textit{as if for a purpose}. Whether that shaping is the result of natural selection, neural plasticity, or gradient descent, it is an optimization process resulting in a system that is effective at performing a task or function, as if it were \textit{designed} to be so.” If part of that purpose for humans, either explicitly or implicitly, is the recognition, production, and dissemination of the beautiful then we should expect machines to approximate our teloi when trained on the data we produce, similarly binding or ordering their representations towards beauty. Furthermore, because beauty is an ends to human culture, we should expect to find beauty as an ordering principle within the physical reality which constrained our own process of cognitive development - and such it should be found also by self-supervised models attempting to apprehending that same reality.

\section{Addressing the Effect of Training Data Congruence}

\begin{figure*}[!ht]

  \includegraphics[width=1\linewidth]{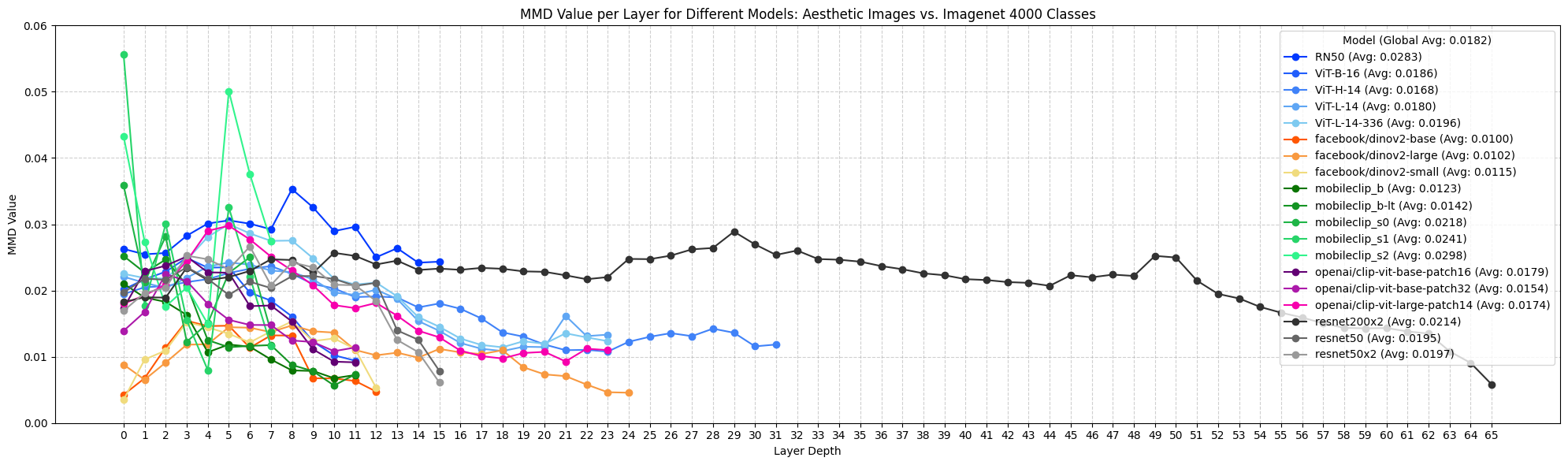}
\raggedright\includegraphics[width=1\linewidth]{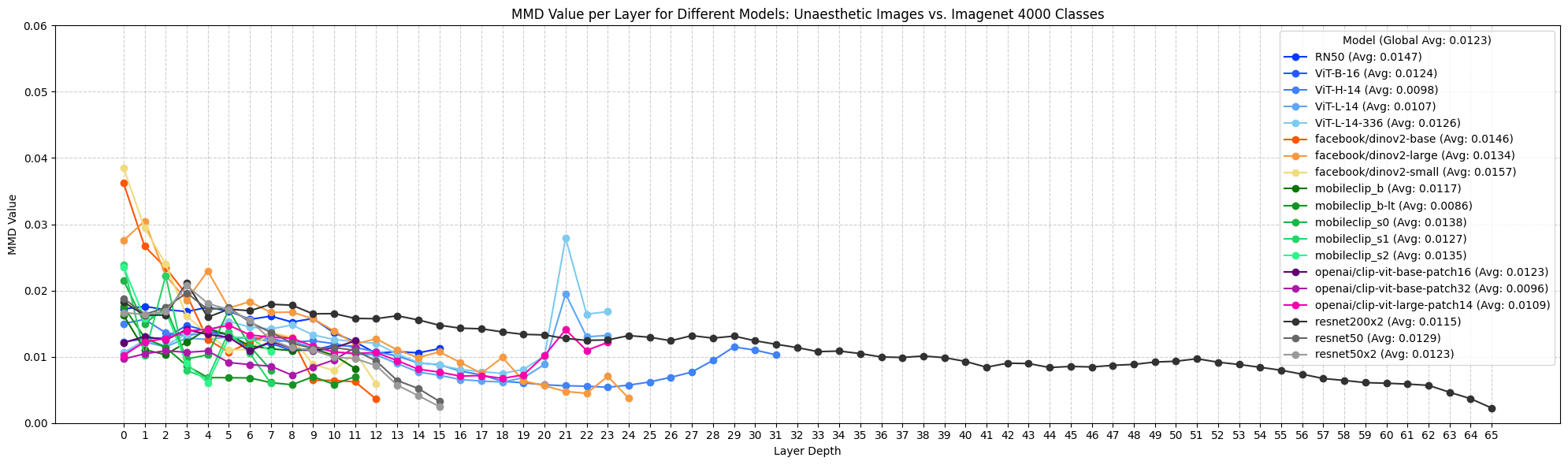}

\caption{Across the majority of layers in our considered models, aesthetic images produced a higher MMD score relative to a 4000 class subset of Imagenet \cite{5206848}, implying that the aesthetic images are not substantially more in domain than the unaesthetic images. The increase in cross-model alignment is thus unlikely to be the result of cultural familiarity with aesthetic images.}

\label{fig:mmd}

\end{figure*}

An underlying claim of post-structuralism is that the relationships between semantic concepts or signs are not fixed, but instead arbitrarily constructed without reference to a concrete reality. Following this conception, an immediate critique of our support for representational objectivity, particularly as it relates to beauty, is that this effect is created by a form of cultural congruence between the model's training data and the aesthetic testing data. In the rote formulation of this argument, this contains a claim of eurocentrism or colonial imposition - that the primarily Western training data of deep learning models forces them to learn a Western ideal of beauty which can therefore be safely ignored as subjective at best, racist at worst. This claim holds less weight when leveled against models trained via self-supervised learning because they receive no human annotation or description in training, so even this ``eurocentric” conception of beauty would have to exist in some real sense within the image image data for it to have been learned and thus be measurable through cross-model representation convergence. Additionally, in Figure \ref{fig:intra-inter} we demonstrated this behavior in ChineseCLIP \cite{yang2023chineseclipcontrastivevisionlanguage}; a model trained almost entirely with data from China - hardly a Western nation by any conception, and one with long standing artistic and aesthetic traditions predating extensive contact with Western nations. But this critique is worth answering more methodically - does an ``in domain” effect increase representation convergence? 

Though none of our tested models mention AVA as part of their training data, and VICReg was trained solely on Imagenet, there is a possibility of data contamination through web crawl data. More likely, the training datasets might simply be similar to the AVA dataset - luckily, we can test this using maximum mean discrepancy (MMD) \cite{JMLR:v13:gretton12a}, a common tool for measuring how out-of-domain a model finds a particular set of input data to be. In Figure \ref{fig:mmd}, we present an MMD analysis of the layer-wise representations of both unaesthetic and aesthetic AVA images relative to embeddings from a four thousand class subset of Imagenet - a dataset which is clearly in-domain both in terms of its Western web crawl origin, its pipeline transformation, and its importance as a training and evaluation dataset. It is clear that both aesthetic and unaesthetic image sets are not perfectly in-domain, but in fact that unaesthetic images are actually more in domain than aesthetic. This substantially weakens claims that aesthetic-mediated representation convergence is created by a habituated understanding of images corresponding to a Western conception of beauty.

\subsection{Semantic Tightness as a Driver of Convergence}

\begin{figure*}[!ht]

  \includegraphics[width=1\linewidth]{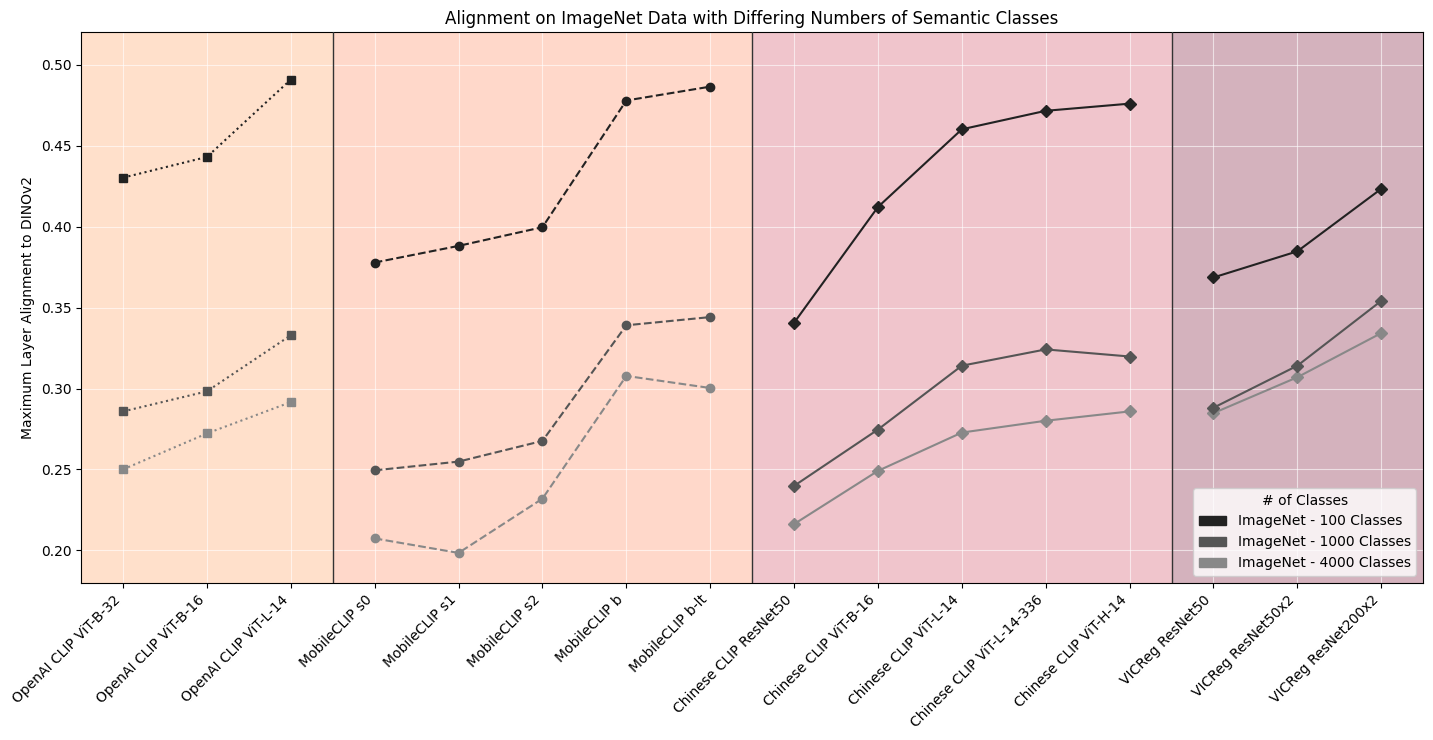}

   \caption{Representational convergence measured on equal size subsets of the Imagenet dataset, delineated by the number of represented classes and agnostic to image aesthetics. }

\label{fig:imagenet}

\end{figure*}

An underlying assumption of this paper is that the inherent semantic coherence of a testing dataset increases measured convergence, though we agree that model ability increases cross-model convergence on identical testing data \cite{ciernik2025objective}. In other words, if a dataset has tight semantic clusters within it that orders the data in question, we would expect cross-model convergence to increase because of the greater amount of mutual information ordering the realized embedding spaces. In order to substantiate this view, we perform an analysis on Imagenet by taking equal size but increasingéy semantically tighter subsets of non-overlapping classes and subjecting them to cross-model convergence analysis. We present these results in Figure \ref{fig:imagenet}, and find that the data which had a smaller number of represented classes achieved higher levels of convergence. 

\begin{figure}[!ht]

  \includegraphics[width=1\linewidth]{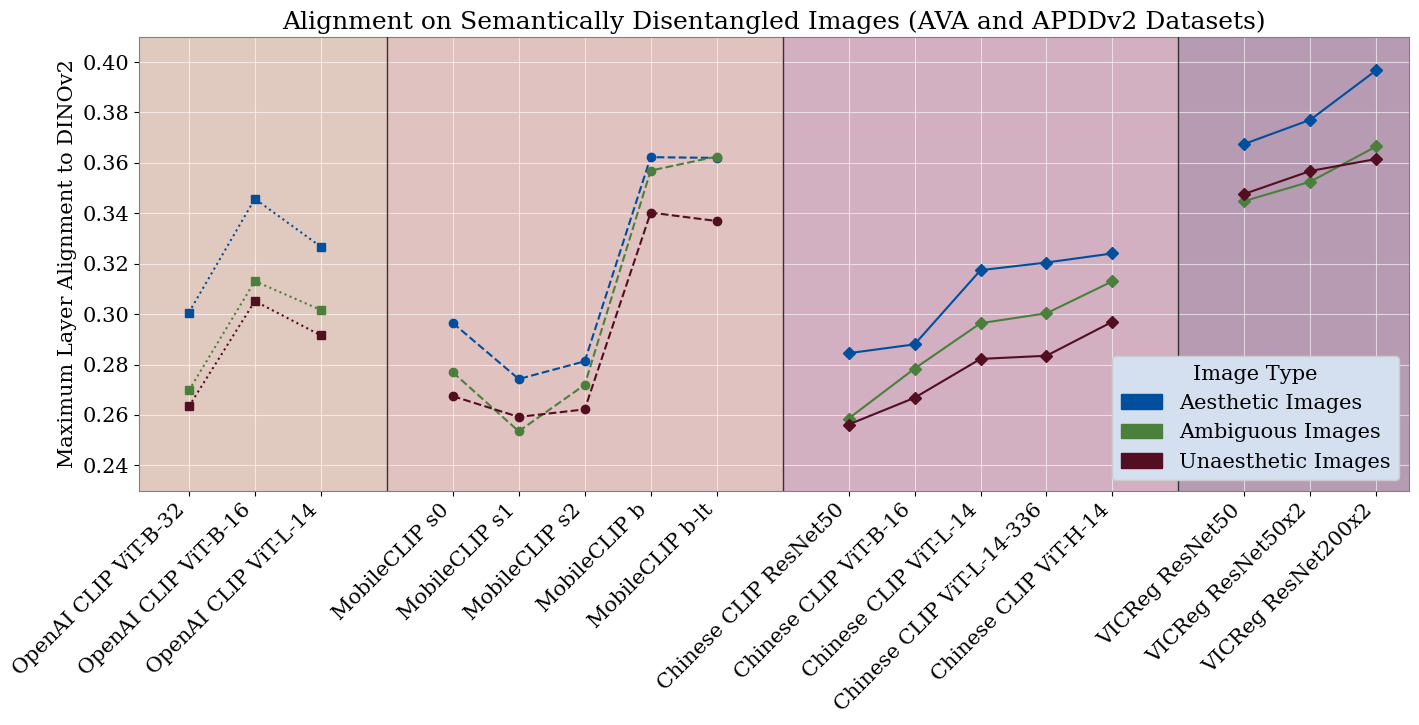}

   \caption{Cross-model alignment measured on international art and image aesthetic datasets AVA and APDDv2 The input data has also been semantically disentangled by ensuring there is only one image per predicted class, thus removing any image subject cluster tightness advantage from the aesthetic image subset. }

\label{fig:disentangle}

\end{figure}

When referring to aesthetic categories, we want to control for semantic clusters as strongly as possible in order to isolate the effect of aesthetic stratification. Though artistic subject or image content is a valid component of aesthetics, we do not want to measure whether the aesthetic section of our dataset simply contains a smaller diversity of subject matter relative to the unaesthetic subset. In order to control for this and for the earlier eurocentric critique, we first introduce the APDDv2 \cite{jin2024apddv2aestheticspaintingsdrawings} artistic aesthetics dataset, which was constructed from primarily Asian artwork and was rated entirely by Chinese nationals. Secondly, we utilize a Torchvision  \cite{torchvision2016} standard Resnet \cite{He_2016_CVPR} model trained on Imagenet (1K) for closed-set classification to predict labels for every image in our combined datasets. With these labels, we enforce that within each aesthetic or unaesthetic subset there can only be one representative of each class - thus controlling for semantic clusters. This leaves use with a balanced dataset of human-rated images from both the East and West, which has also been semantically disentangled in order to isolate the content of aesthetics. In Figure \ref{fig:disentangle} we show that, despite this substantially more difficult setting, we recover the same aesthetic stratification in cross-model convergence. The persistence of this effect should motivate further research in this direction to  isolate the specific drivers of aesthetic convergence in representation space.

\section{Representational Projection through the Artistic Act}

Influential philosopher of art Arthur C. Danto succinctly defined a work of art as ``embodied meaning” \cite{Danto1981-transfig}. For Danto, the fundamental characteristic of the work of art was unrelated to aesthetics, formal rendering category (such as the stereotypical painting or marble sculpture), or other intrinsic foundation, but purely based on the representation of semantic meaning in the physical world and the recognition of that semantic meaning by the institutional art world. Assuming his semantic meanings map cleanly to the universal semantic representations we have been discussing, then Danto's definition of art is predicated on the projection of these representations as perceived, filtered, and rendered by the artists and art institutions. Though Danto contested that semantic projection extended to `mere objects' because they lacked an `aboutness', this perspective could be challenged by the success in broad semantic understanding of all manner of objects, through various modalities, by deep learning systems. This conception of the artistic act is remarkably similar to that of classical philosopher Plotinus: ``the stone thus brought under the artist’s hand to the beauty of form is beautiful not as stone ... but in virtue of the form or idea introduced by the art. This form is not in the material; it is in the designer before ever it enters the stone; and the artificer holds it ... by his participation in his art” \cite{Gerson_2017}. Though Plotinus clearly has a more Platonic conception of form because he does not incorporate the cultural reception of the work of art as Danto stresses, both center the human phenomenological lens as participatory and projective of form. 
\mycomment{
Finally, we also see the same conception of artistic projection of an inner ideal in the `concetto' of the Renaissance and artists, here presented by Michelangelo: ``Nothing the best of artists can conceive but lies, potential, in a block of stone, superfluous matter round it. The hand alone can free it that has intelligence for guide” .
}

An underexplored factor in discussions surrounding the universal representation hypothesis is the degree to which deep learning training data is filtered through the human phenomenological lens. Though we do not dispute that there are underlying physical constraints at play, it is important to recognize that there is no part of a deep learning system which is not distinctly shaped by human intention and perspective. This is true of their architectures, which have been explicitly inspired by the human conception of our own brains and the sort of operations they perform, to the human-mimicking tasks which are set before them, to the human captured, created, and curated data on which they are trained. Bringing this perspective to bear in light of our discussion of universal representations serves to center the human artistic, aesthetic, and creative impulses as a primary driver for making the world machine intelligible. It is through human creativity, whether as simple as labeling an image with a text description or as complex as the construction of a Gothic cathedral, that we distill and project what we find in nature; we imbue the created object with a refined form of semantic meaning which substantially simplifies the path of a deep learning system to likewise apprehending the universal.

\section{The Place of the Machine in Cultivating the Beautiful}

The source of the ability for deep learning systems to understand artistically relevant concepts like beauty has a substantial effect on the way these models will be used and is of particular interest to artists who engage with these systems. If we understand the ground of beauty to be purely subjective and structurally arbitrary, then there is little for deep learning systems to add to the conversation outside of facilitating the production of particular subjective aesthetics. If beauty is purely objective and disconnected from the human phenomenological experience, then we should expect that deep learning systems exposed to more data should eventually become the optimal technique for producing beautiful objects to the exclusion of the now redundant human artist. However, we claim that there is an objective (material, \textit{hulê}) basis for beauty which is subjectively apprehended (phenomenological, \textit{morphê}) before returning through the creative act as unified hylomorphic form. This conception centers the perception and work of humans, but also allows for deep learning systems to contribute meaningfully to the conversation around beauty by `perceiving' these representations at scale, thus refining, interrogating, and expanding human conceptions of beauty. We are hopeful that this symbiotic framework can help to resolve some of the tensions between human creators and machines, allowing us to more easily co-cultivate the beautiful.

\printbibliography

\end{document}